\documentstyle[12pt]{article}
\begin{document}
\begin{titlepage}
\begin{flushright}
Alberta Thy-08-95\\ February, 1995

\end{flushright}
\vskip 10mm

\begin{center}
{\Large \bf  Nonfactorization in Hadronic Two-body Cabibbo-favored
decays of $D^0$ and $D^+$  }\\
\vskip 5mm
A. N. Kamal and A. B. Santra\footnote{on leave of absence from
Nuclear Physics Division, Bhabha Atomic Research Centre, Bombay
400085,  India.}\\
{\em Theoretical Physics Institute and Department of Physics,\\
University of Alberta,
Edmonton, Alberta T6G 2J1, Canada.}\\
\vskip 1mm
T. Uppal\\
{\em Department of Physics, Fairfield University\\
Fairfield, Connecticut 06430, U.S.A.\\}
and\\
R. C. Verma\\
{\em Department of Physics,Panjab University\\
Chandigarh 160014, India.}
\end{center}
\vskip 5mm

\begin{abstract}
With the inclusion of nonfactorized amplitudes in a scheme with
$N_c=3$, we have studied Cabibbo-favored decays of $D^0$ and $D^+$
into  two-body hadronic states involving two isospins in the final
state. We have shown that it is possible to understand the measured
branching ratios and determined the sizes and signs of nonfactorized
amplitudes required.
\end{abstract}
PACS index: 13.25 Hw, 14.40. Nd
\end{titlepage}

\begin{center}
{\bf I. Introduction}
\end{center}

In recent past there has been a growing interest  [1-7] in exploring
the role played  by nonfactorized terms in the hadronic decays of
charmed and beauty mesons. Ref. \cite{shf} and \cite{khod} have
endeavored to calculate the nonfactorized contribution to two-body
hadronic decays of the B meson. These calculations lend support to
the $N_c \rightarrow \infty$ rule in two-body hadronic B decays.
Experimentally however, the evidence in support \cite{nbt} of the
$N_c \rightarrow \infty$ rule which appeared to be there in the
earlier B-decay data has since weakened \cite{alm} and the sign of
the phenomenological parameter $a_2$ appears to be positive
\cite{alm} contrary to the prediction of the $N_c \rightarrow \infty$
rule.

More recently, the view that the phenomenological parameters $a_1$
and $a_2$ are effective and process-dependent has been pursued
further [3-7]. The effective $a_1$ and $a_2$, evaluated with $N_c =
3$, depend on the nonfactorized contribution. In particular  it was
shown in Ref. \cite{ks1}  how the conundrum of the failure \cite{gkm}
of all popular models to explain the longitudinal polarization
fraction in $B^0 \rightarrow \psi \bar{K}^{*0}$ could be resolved in
a scheme that uses $N_c = 3$ but allows a small nonfactorized
amplitude. This idea was carried over to the charm sector in Ref.
\cite{ks2} where it was shown that with $N_c = 3$ allowing
nonfactorized terms somewhat larger than in B decays (by
nonfactorized terms `large' or `small' we mean: in relation to
factorized terms),  one could understand data in $D^+_s \rightarrow
\phi \pi^+, \phi \rho^+$ and $\phi l^+ \nu_l$ decays. The
introduction and description of nonfactorized terms is purely
phenomenological in Refs. \cite{ks1,ks2} as is also the case in
\cite{chng, srs,hyc}. No attempt is made to calculate the
nonfactorized terms but, rather, the emphasis is to glean some
systematic behavior of  these terms so that more can be learned about
them in future.

With this objective  we have studied  those hadronic two-body
Cabibbo-favored decays of $D^0, D^+$  mesons that involve two
isospins in the final state in $N_c = 3$ scheme.  These decays are:
$D \rightarrow \bar{K} \pi$, $ \bar{K}^* \pi$, $\bar{K} \rho$,
$\bar{K}a_1$ and $\bar{K}^* \rho$. By fitting data, we have
calculated the size and the sign of the nonfactorized term in each
decay. Annihilation terms wherever permitted have been neglected in
$D^0$ decays due to the smallness of $a_2$ ($=\;C_2 + {C_1 \over
N_c}$) for $N_c =3$. We have included final-state interaction phases
wherever they have been determined experimentally, examples are the
decays  $D^0 \rightarrow K^- \pi^+$, $\bar{K}^0 \pi^0$,  $D^0
\rightarrow K^- \rho^+$, $\bar{K}^0 \rho^0$ and $D^0 \rightarrow
K^{*-} \pi^+$, $\bar{K}^{*0} \pi^0$.   However, we have neglected
inelastic final state interactions due to the ignorance of the
rescattering parameters to be used in such an analysis.

For decays involving a single Lorentz scalar structure, such as $D
\rightarrow \bar{K} \pi$, $ \bar{K}^* \pi$, $\bar{K} \rho$ and
$\bar{K}a_1$, one can extract effective $a_1$ and $a_2$ which we show
to be process-dependent. We also argue that color-suppressed decays
are more likely to reveal presence or otherwise of nonfactorized
effects.

This paper is organized as follows: Section II contains the
conventions and definitions used throughout.  We discuss the decays
$D \rightarrow \bar{K} \pi$  in Section III,  $D \rightarrow
\bar{K}^* \pi$, $\bar{K} \rho$,  $\bar{K}a_1$  in Section IV  and $D
\rightarrow \bar{K}^* \rho$  in Section V. The results are discussed
in Section VI.

\vskip 1cm
\begin{center}
{\bf II.  Definitions}
\end{center}

The effective Hamiltonian for Cabibbo-favored  hadronic charm decays
is given by
\begin{eqnarray}
H_w = \tilde{G}_F \left\{ C_1\left( \bar{u}d \right) \left( \bar{s}c
\right) + C_2 \left( \bar{u}c \right)\left( \bar{s}d \right) \right\}
\;,
\end{eqnarray}
where $\tilde{G}_F = {G_F\over \sqrt{2} } V_{cs}V_{ud}^*$ and
$(\bar{u}d)$ etc. represent color-singlet (V-A) Dirac currents.
$C_1$ and $C_2$ are the Wilson coefficients for which we adopt the
following values,
\begin{eqnarray}
C_1 = 1.26 \pm 0.04\;, \qquad C_2 = -0.51 \pm 0.05 \;.
\end{eqnarray}
The central values of $C_1$ and $C_2$ are taken from Ref. \cite{nbt}
and the errors are ours.

Fierz transforming the product of two Dirac currents of eqn.(1) in
$N_c$-color-space, we get,
\begin{eqnarray}
\left( \bar{u}c \right) \left( \bar{s} d \right) = {1 \over N_c}
\left( \bar{u} d \right) \left( \bar{s} c \right) + {1 \over 2}
\sum_{a =1}^{8}{\left( \bar{u} \lambda^a d \right) \left( \bar{s}
\lambda^a  c \right) } \; ,
\end{eqnarray}
and an analogous relation for $\left( \bar{u} d \right) \left(
\bar{s} c \right)$, where $\lambda^a$ are the Gell-Mann matrices.
Using eqn.(3) and its analogue,  we reduce  the effective Hamiltonian
of eqn.(1)  to the following forms
\begin{eqnarray}
H_w^{CF}&=&\tilde{G}_F\left\{ a_1\left( \bar{u}d \right) \left(
\bar{s}c \right) + C_2 H_w^8\right\} \;, \\
\hbox{and}\qquad \qquad  H_w^{CS}&=&\tilde{G}_F\left\{ a_2 \left(
\bar{u}c \right)\left( \bar{s}d \right) + C_1\tilde{H}_w^8\right\}
\;,
\end{eqnarray}
to describe color-favored (CF) and color-suppressed (CS) decays
respectively. The matrix elements of the first terms in  (4) and (5)
are  expected to be dominated by factorized contributions; any
nonfactorized part  arising from them is parametrized as detailed in
the text.     The second terms, $H_w^{(8)}     ( \equiv {1 \over 2}
\sum{(\bar{u}\lambda^a d)(\bar{s}\lambda^a c)} )$     and
$\tilde{H}_w^{(8)}     ( \equiv {1 \over 2} \sum{(\bar{u}\lambda^a
c)(\bar{s}\lambda^a d)})     $,   involving color-octet currents
aenerate  nonfactorized contributions. We have  defined here for $N_c
= 3$,
\begin{eqnarray}
a_1 = C_1 + {C_2 \over 3} = 1.09 \pm 0.04\;, \qquad
a_2 = C_2 + {C_1 \over 3} = -0.09 \pm 0.05\;.
\end{eqnarray}
It should be obvious from (4) and (5) that nonfactorized effects are
more likely to manifest themselves in color-suppressed decays than in
color-favored decays due to the fact that $C_1$ is much larger than
$a_2$ in magnitude.

Further, in calculating the factorized amplitudes, we use the
following matrix elements \cite{nbt, bsw} for the weak vector
($j_\mu^V$) and axial vector ($j_\mu^A$) currents between the vacuum
and the pseudoscalar (P), vector (V) and axial vector (A) states

\begin{eqnarray}
\left\langle V(p,\varepsilon) |j_\mu^V|
0\right\rangle&=&\varepsilon_\mu^* m_V f_V \;,\nonumber \\
\left\langle A(p,\varepsilon) |j_\mu^A|
0\right\rangle&=&\varepsilon_\mu^* m_A f_A \;,\nonumber \\
\left\langle P(p) |j_\mu^A| 0\right\rangle&=&- i f_P p_\mu \; ,
\end{eqnarray}
and the  form factors for the transition of a pseudoscalar meson (M)
to pseudoscalar (P) and vector (V) mesons,
\begin{eqnarray}
\left\langle P(p^\prime)|j^A_\mu|M(p) \right\rangle=\left\{ \left( p
+ p^\prime \right)_\mu - { m_M^2 - m_P^2\over q^2} q_\mu \right\}
F_1^{MP}\left( q^2 \right) + { m_M^2 - m_P^2\over q^2} q_\mu
F_0^{MP}\left( q^2 \right)\; ,
\end{eqnarray}
\begin{eqnarray}
\left\langle V(p^\prime,  \varepsilon ) | j^V_\mu - j^A_\mu | M(p)
\right\rangle&=&i \left\{ (m_M + m_V) \varepsilon_\mu^* A_1^{MV}(q^2)
- {\varepsilon^*.q \over m_M + m_V} ( p + p^\prime)_\mu A_2^{MV}
(q^2) \right. \nonumber \\   &-& \left. 2m_V {\varepsilon^*.q \over
q^2} q_\mu \left( A_3^{MV}(q^2)  - A_0^{MV}(q^2) \right) \right\}
\nonumber \\
 &+&{2 \over m_M + m_V} \varepsilon_{\mu \nu \rho \sigma}
\varepsilon^{*\nu} p^\rho p^{\prime \sigma }V^{MV}(q^2) \; ,
\end{eqnarray}
where $q_\mu = (p - p^\prime)_\mu$.  In addition, the following
constraint applies at all $q^2$,
\begin{eqnarray}
2m_V A_3^{MV}(q^2) = (m_M + m_V) A_1^{MV}(q^2) - (m_M - m_V)
A_2^{MV}(q^2) \; .
\end{eqnarray}
 The following relation is needed to cancel the poles at $q^2 = 0$,
\begin{eqnarray}
A_0^{MV}(0) = A_3^{MV}(0).
\end{eqnarray}

For an axial vector meson,A, we define analogously to (9),
\begin{eqnarray}
\left\langle A(p^\prime,  \varepsilon ) | j^V_\mu - j^A_\mu | M(p)
\right\rangle&=&-i \left\{ (m_M + m_A) \varepsilon_\mu^*
V_1^{MA}(q^2)  - {\varepsilon^*.q \over m_M + m_A} ( p +
p^\prime)_\mu V_2^{MA} (q^2) \right. \nonumber \\   &-& \left. 2m_A
{\varepsilon^*.q \over q^2} q_\mu \left( V_3^{MA}(q^2)  -
V_0^{MA}(q^2) \right) \right\} \nonumber \\
 &-&{2 \over m_M + m_A} \varepsilon_{\mu \nu \rho \sigma}
\varepsilon^{*\nu} p^\rho p^{\prime \sigma }A^{MA}(q^2) \; ,
\end{eqnarray}
with the same conditions (10) and (11) imposed on $V_i^{MA}(q^2)$.

The branching ratio for $M \rightarrow M_1M_2$, where $M_1$ and $M_2$
are pseudoscalor mesons, is given by

\begin{eqnarray}
B (M \rightarrow M_1M_2) = \tau_M {|\vec{p}| \over 8 \pi m_M^2} |A(M
\rightarrow M_1M_2)|^2 \;,
\end{eqnarray}

and that for $M \rightarrow V_1V_2$, where $V_1$ and $V_2$ are vector
mesons, is written as
\begin{eqnarray}
B (M \rightarrow V_1V_2) = \tau_M {|\vec{p}| \over 8 \pi m_M^2}
\sum_{\lambda}^{}{|A(M \rightarrow V_1V_2)_{\lambda \lambda}|^2} \;,
\end{eqnarray}

where $|\vec{p}| $ is the magnitude of final-state three-momentum in
M-rest frame, $\tau_M$ is the life time of M and A($M \rightarrow
M_1M_2$) etc.  are the decay amplitudes. The branching ratio formula
for $M \rightarrow PV$ decay is the same as (14) with a sum over
polarizations of V.

In the following, we list some of the parameters we have used
throughout this paper:
\begin{eqnarray}
f_\pi&=&130.7 \; \hbox{MeV}, \qquad f_K = 159.8 \;
\hbox{MeV},\nonumber \\
f_\rho&=&212.0 \; \hbox{MeV}, \qquad f_{K^*} = 221.0 \; \hbox{MeV},
\nonumber \\
f_{a_1}&=&212.0 \; \hbox{MeV}, \nonumber \\
V_{cs}&=&0.975 \;, \qquad V_{ud} = 0.975 .
\end{eqnarray}

\vskip  5mm
\begin{center}
{\bf III.  $D \rightarrow  P_1 P_2$}\\
\vskip  2mm
{\bf A. $D^0 \rightarrow K^- \pi^+ \; , \; \bar{K}^0 \pi^0 $ and
$D^+ \rightarrow \bar{K}^0 \pi^+ $ }
\end{center}
\vskip  2mm

To illustrate our method we write,  using eqn. (4) for the effective
Hamiltonian, the decay amplitude of  $D^0 \rightarrow  K^- \pi^+$ as,
\begin{eqnarray}
A(D^0 \rightarrow K^- \pi^+) = \tilde{G}_F  \left\{  a_1 \left\langle
K^- \pi^+|(\bar{s}c)(\bar{u}d)| D^0 \right\rangle  + C_2 \left\langle
K^- \pi^+ |H_w^{(8)}| D^0\right\rangle \right\} \; .
\end{eqnarray}
We write the first term as a sum of a factorized and a nonfactorized
part,
\begin{eqnarray}
\left\langle K^- \pi^+ \bigl| (\bar{s}c)(\bar{u}d) \bigr|
D^0\right\rangle&=&\left\langle  \pi^+ \bigl| (\bar{u}d) \bigr| 0
\right\rangle \left\langle K^-  \bigl| (\bar{s}c)\bigr|
D^0\right\rangle  +    \left\langle \pi^+ K^-|(\bar{s}c)(\bar{u}d)|
D^0 \right\rangle^{nf}   \nonumber \\
&=& -i f_\pi (m_D^2 - m_K^2) (F_0^{DK}(m_\pi^2) +
F_0^{(1)nf}(m_\pi^2))\; ,
\end{eqnarray}
where we have defined the nonfactorized matrix element of the product
of the color-singlet currents $(\bar{s}c)(\bar{u}d)$ as
\begin{eqnarray}
\left\langle K^-\pi^+|(\bar{s}c)(\bar{u}d)|D^0 \right\rangle^{nf}
\equiv -if_\pi (m_D^2 - m_K^2) F_0^{(1)nf}(m_\pi^2)\; .
\end{eqnarray}
For  the second term in (16) we write,
\begin{eqnarray}
\left\langle K^- \pi^+ \bigl| H_w^{(8)} \bigr| D^0\right\rangle = -i
f_\pi (m_D^2 - m_K^2) F_0^{(8)nf}(m_\pi^2) \; .
\end{eqnarray}
The decay amplitude of eqn. (16) is  then  written in the form,

\begin{eqnarray}
A(D^0 \rightarrow K^- \pi^+ ) = -i \tilde{G}_F   (a_1 ^{eff})_{K \pi}
f_\pi \left( m_D^2 - m_K^2 \right)  F_0^{DK}(m_\pi^2) \; ,
\end{eqnarray}
where,
\begin{eqnarray}
\left( a_1^{eff} \right)_{K \pi} = a_1 \left( 1 +
{F_0^{(1)nf}(m_\pi^2) \over F_0^{DK}(m_\pi^2)} + {C_2 \over a_1}
{F_0^{(8)nf}(m_\pi^2) \over F_0^{DK}(m_\pi^2)} \right) \; .
\end{eqnarray}
This defines  a process-dependent effective $a_1$.  We shall see
that it is possible to do so for all decays involving a single
Lorentz scalar structure.  We notice also that as the coefficient
$C_2 / a_1$ ($\approx -0.47$) is smaller than unity,   the effect of
the nonfactorized amplitude is suppressed relative to the factorized
amplitude  in color-favored decays.  For the same reason, the
nonfactorized term proportional to $F^{(1)nf}_0$ could compete
favorably with $F^{(8)nf}_0$.

The decay amplitude for the color-suppressed decay $D^0 \rightarrow
\bar{K}^0 \pi^0$ by following an analogous procedure is given by,
\begin{eqnarray}
A(D^0 \rightarrow \bar{K}^0 \pi^0 ) = -i {\tilde{G}_F \over \sqrt{2}}
(a_2 ^{eff})_{K \pi} f_K \left( m_D^2 - m_\pi^2 \right)
F_0^{D\pi}(m_K^2) \; ,
\end{eqnarray}
where,
\begin{eqnarray}
\left( a_2^{eff} \right)_{K \pi} = a_2 \left( 1 +  {C_1 \over a_2}
{\tilde{F}_0^{(8)nf}(m_K^2) \over F_0^{D\pi}(m_K^2)} \right) \; .
\end{eqnarray}

In writing (22) we have used,
\begin{eqnarray}
\left\langle \bar{K}^0 \pi^0 | (\bar{u}c)(\bar{s}d) | D^0
\right\rangle &=& \left\langle \bar{K}^0|(\bar{s}d)| 0\right\rangle
\left\langle \pi^0|(\bar{u}c)| D^0\right\rangle + \left\langle
\bar{K}^0 \pi^0 | (\bar{u}c)(\bar{s}d) | D^0 \right\rangle^{nf} \; ,
\nonumber \\
&\approx& -i{ f_K\over \sqrt{2}} (m_D^2 - m_\pi^2) F_0^{D\pi}(m_K^2)
\;,
\end{eqnarray}
and
\begin{eqnarray}
\left\langle \bar{K}^0 \pi^0| \tilde{H}_w^{(8)}| D^0\right\rangle =
-i {f_K \over \sqrt{2}} (m_D^2 - m_\pi^2)
\tilde{F}_0^{(8)nf}(m_K^2)\;.
\end{eqnarray}

Now, as  ${C_1 \over a_2}$  in eqn. (23) is large ($\approx -14$),
the nonfactorized contribution  arising from $\tilde{H}_w^{(8)}$ is
greatly enhanced. In contrast,  any possible nonfactorized effects in
(24) are suppressed  due to the smallness of $a_2$.  For this reason
we have neglected the nonfactorized contribution in (24).

The amplitude for  $D^+ \rightarrow \bar{K}^0 \pi^+$  decay is
obtained from eqns. (20) and (22) via the isospin sum rule
\begin{eqnarray}
A(D^+ \rightarrow  \bar{K}^0 \pi^+) = A(D^0 \rightarrow  {K}^- \pi^+)
+ \sqrt{2} A(D^0 \rightarrow  \bar{K}^0 \pi^0)\;.
\end{eqnarray}

In terms of isospin amplitudes $A_{1/2}$ and  $A_{3/2}$ and the
final-state interaction (fsi) phases,
\begin{eqnarray}
A(D^0 \rightarrow K^- \pi^+)&=&{1 \over \sqrt{3}}\left( A_{3/2}exp(i
\delta_{3/2}) + \sqrt{2}A_{1/2}exp(i \delta_{1/2}) \right) \nonumber
\\
A(D^0 \rightarrow \bar{K}^0 \pi^0)&=&{1 \over \sqrt{3}}\left(
\sqrt{2}A_{3/2}exp(i \delta_{3/2}) - A_{1/2}exp(i \delta_{1/2})
\right) \nonumber \\
A(D^+ \rightarrow \bar{K}^0 \pi^+)&= &\sqrt{3} A_{3/2}exp(i
\delta_{3/2})
\end{eqnarray}
The relative phase is known \cite{kp} to be
\begin{eqnarray}
\delta^{\bar{K}\pi}_{1/2} - \delta^{\bar{K}\pi}_{3/2} = (86 \pm 8)^0
\end{eqnarray}

We determine $A_{1/2}$ and $A_{3/2}$ by equating eqns. (20) and (22)
to eqn. (27) with the phases $\delta_{1/2}$ and $\delta_{3/2}$ set
equal to zero; and then reinstate the phases  to calculate  the
branching ratios  from eqn.(13). This procedure is equivalent to
assuming that the effect of fsi in this mode is simply to rotate the
isospin amplitudes without effecting their magnitudes.  For the form
factors we have used the following normalizations at $q^2 = 0$,
\begin{eqnarray}
F_0^{DK}(0)&=&0.77 \pm 0.04\;, \qquad \cite{wthrl} \nonumber \\
F_0^{D\pi}(0)&=&0.83 \pm 0.08\;. \qquad \cite{kp,chau}
\end{eqnarray}

In practice we have used only the central values of these form
factors and extrapolated $F_0^{DK}(q^2)$ and $F_0^{D\pi}(q^2)$ as
monopoles with $0^+$ pole masses of 2.01 and  2.47 GeV respectively
as in Ref. \cite{bsw}.  As these form factors are needed at  a
relatively small  $q^2$ (=$m^2_\pi$ or $m^2_K$), the results are not
very sensitive to the manner of extrapolation.

The results are summarized below:  Defining

\begin{eqnarray}
 \chi_{K\pi}  \equiv {F_0^{(8)nf} \over F_0^{DK}(m_\pi^2)} + {a_1
\over C_2} {F_0^{(1)nf} \over F_0^{DK}(m_\pi^2)}    \qquad \hbox{and}
\qquad  \xi_{K\pi}  \equiv {\tilde{F}_0^{(8)nf} \over
F_0^{D\pi}(m_K^2)}
\end{eqnarray}
we  get agreement with the data for nonzero $\chi_{K\pi} $ and
$\xi_{K\pi} $ only with the fsi relative phase lying in the following
range,
\begin{eqnarray}
77^0 \leq \delta^{\bar{K}\pi}_{1/2} - \delta^{\bar{K}\pi}_{3/2}
\leq103^0 \;.
\end{eqnarray}
Taking $\delta^{\bar{K}\pi}_{1/2} - \delta^{\bar{K}\pi}_{3/2} =
90^0$, we determine the  allowed ranges of $\chi_{K\pi}$ and
$\xi_{K\pi}$ to be,
\begin{eqnarray}
-0.16 \leq \chi_{K\pi } \leq-0.08\;, \qquad -0.29\leq \xi_{K\pi}\leq
-0.26 \;.
\end{eqnarray}
These  ranges of $\chi_{K\pi}$ and $\xi_{K\pi}$ translate into the
following ranges of $(a_1^{eff})_{K \pi}$ and $(a_2^{eff})_{K \pi}$,
\begin{eqnarray}
1.13 \leq(a_1^{eff})_{K \pi} \leq1.17 \;, \qquad \hbox{and} \qquad
-0.46 \leq (a_2^{eff})_{K \pi} \leq-0.42\;.
\end{eqnarray}
In particular, with $\delta^{\bar{K}\pi}_{1/2} -
\delta^{\bar{K}\pi}_{3/2} = 90^0$,  $\chi_{K\pi} = -0.12$ and
$\xi_{K\pi} =-0.27$, we obtain,
\begin{eqnarray}
B(D^0 \rightarrow K^-\pi^+) = 3.99 \% \;(\hbox{expt.}\cite{pdg}:
(4.01 \pm 0.14)\% )\nonumber \\
B(D^0 \rightarrow \bar{K}^0\pi^0) = 2.17 \%
\;(\hbox{expt.}\cite{pdg}: (2.05 \pm 0.26)\% )\nonumber \\
B(D^+ \rightarrow \bar{K}^0\pi^+) = 2.76 \%
\;(\hbox{expt.}\cite{pdg}: (2.74 \pm 0.29)\% )
\end{eqnarray}

Clearly,  with nonfactorized contribution proportionately larger in
$D^0 \rightarrow \bar{K}^0\pi^0$ than in $D^0 \rightarrow
\bar{K}^-\pi^+$, it is possible to understand data in a scheme with
$N_c = 3$. The amount of nonfactorized amplitude needed is reasonably
small. We wish to emphasize that an annihilation term, if present,
would be much suppressed in our description since such a term would
be proportional to $a_2$ which in $N_c=3$ scheme is only $\approx
-0.09$. Past estimates \cite{kp,chau} of allowed annihilation terms
were based on the $N_c \rightarrow \infty$ value of $a_2=-0.51$.  We
shall return to a discussion of our numerical estimates of
$\chi_{K\pi}$ and $\xi_{K\pi}$ (equivalently $(a_1^{eff})_{K\pi} $
and  $(a_2^{eff})_{K\pi}$   ) in  Section VI.

\vskip  5mm
\begin{center}
{\bf IV.  $D \rightarrow  P_1 V_1$}\\
\vskip  2mm
{\bf A. $D^0 \rightarrow K^{*-} \pi^+ \; , \; \bar{K}^{*0} \pi^0 $
and  $D^+ \rightarrow \bar{K}^{*0} \pi^+ $ }
\end{center}
\vskip  2mm

Using the definitions introduced in Section II and the method of
calculation detailed for $D \rightarrow \bar{K} \pi$ decays,  the
amplitudes for the decays $D^0 \rightarrow K^*\pi $ are given by
\begin{eqnarray}
A(D^0 \rightarrow K^{*-} \pi^+ )&=&2 \tilde{G}_F  f_\pi  m_{K^*}
A_0^{DK^*}(m_\pi^2)  (\varepsilon^*.p_D) (a_1 ^{eff})_{K^* \pi} \; ,
\nonumber \\
A(D^0 \rightarrow \bar{K}^{*0} \pi^0 )&=&\sqrt{2} \tilde{G}_F
f_{K^*} m_{K^*} F_1^{D\pi}(m_{K^*}^2)  (\varepsilon^*.p_D) (a_2
^{eff})_{K^* \pi} \; ,  \nonumber \\
 A(D^+ \rightarrow \bar{K}^{*0} \pi^+ )&=&A(D^0 \rightarrow K^{*-}
\pi^+ ) + \sqrt{2}A(D^0 \rightarrow \bar{K}^{*0} \pi^0 )\;.
\end{eqnarray}
where
\begin{eqnarray}
\left( a_1^{eff} \right)_{K^* \pi}&=&a_1 \left( 1 + {A_0^{(1)nf}
\over A_0^{DK^*}(m_\pi^2)} +  {C_2 \over a_1} {A_0^{(8)nf} \over
A_0^{DK^*}(m_\pi^2)} \right) \; ,\nonumber \\
\left( a_2^{eff} \right)_{K^* \pi}&=&a_2 \left( 1 +  {C_1 \over a_2}
{\tilde{F}_1^{(8)nf}\over F_1^{D\pi}(m_{K^*}^2)} \right) \; .
\end{eqnarray}
In (35) and  (36),  in addition to (8) and (9), we have used the
following definitions
\begin{eqnarray}
\left\langle K^{*-} \pi^+| (\bar{s}c) (\bar{u}d)|D^0
\right\rangle^{nf}&=&2 \tilde{G}_F f_\pi m_{K^*} A_0^{(1)nf}(m_\pi^2)
(\varepsilon^*.p_D) \; ,\nonumber \\
\left\langle K^{*-} \pi^+| H_w^{(8)} |D^0 \right\rangle&=&2
\tilde{G}_F f_\pi m_{K^*} A_0^{(8)nf}(m_\pi^2)  (\varepsilon^*.p_D)
\; ,\nonumber \\
\hbox{and} \qquad  \left\langle \bar{K}^{*0} \pi^+| \tilde{H}_w^{(8)}
|D^0 \right\rangle&=&\sqrt{2} \tilde{G}_F f_{K^*} m_{K^*}
\tilde{F}_1^{(8)nf}(m_\pi^2)  (\varepsilon^*.p_D) \; .
\end{eqnarray}

It is known \cite{kp} that fsi phases in this decay are large,
$\delta^{K^*\pi}_{1/2} - \delta^{K^* \pi}_{3/2} = (103 \pm 17)^0$.
To take the fsi phases into account  we follow a  procedure similar
to that for  $D \rightarrow K\pi$ decays;  we calculate the isospin
amplitudes by equating the amplitudes in (35) to those in (27) with
phases set equal to zero. Having so determined $A_{1/2}$ and
$A_{3/2}$, we reinstate the phases. For the form factors we have used
the following normalizations at $q^2 = 0$,
\begin{eqnarray}
A_0^{DK^*}(0)&=&0.70 \pm 0.09\;, \qquad \cite{wthrl,kp} \nonumber \\
F_1^{D\pi}(0)&=&0.83 \pm 0.08\;. \qquad \cite{kp,chau}
\end{eqnarray}
In actual calculation we have used  only  the central values and we
have considered  monopole (referred to as BSWI hereafter) as well as
dipole (referred to as BSWII hereafter) forms for the
$q^2$-extrapolation of the  form factors $A_0^{DK^*}(q^2)$ and
$F_1^{D\pi}(q^2) $ with pole masses 2.11 and 1.87 GeV respectively.
The results are shown in Table 1 where data are fitted for
$\delta^{K^*\pi}_{1/2} - \delta^{K^* \pi}_{3/2} = 110^0$,  with
$\chi_{K^*\pi}$ and $\xi_{K^*\pi}$, defined in the Table, in the
ranges indicated.   We point out that non-empty domains of $\chi_{K^*
\pi}$ and $\xi_{K^* \pi}$ were found for   $\delta^{K^*\pi}_{1/2} -
\delta^{K^* \pi}_{3/2}$ lying in the interval  ($54^0$ - $125^0$).
The corresponding ranges of  effective $a_1$ and $a_2$ in BSWI and
BSWII scenarios are given as follows
\begin{eqnarray}
\hbox{BSWI} \qquad 1.77\leq \left( a_1^{eff} \right)_{K^* \pi}
\leq1.92 \;, \qquad -0.52 \leq \left(a_2^{eff} \right)_{K^* \pi}
\leq-0.44 \nonumber \\
\hbox{BSWII} \qquad 1.77\leq \left( a_1^{eff} \right)_{K^* \pi}
\leq1.91\;, \qquad -0.42 \leq \left(a_2^{eff} \right)_{K^* \pi}
\leq-0.35
\end{eqnarray}
A discussion of these results is given in Section VI.

\vskip  5mm
\begin{center}
{\bf B. $D^0 \rightarrow K^{-} \rho^+ \; , \; \bar{K}^{0} \rho^0 $
and  $D^+ \rightarrow \bar{K}^{0} \rho^+ $ }
\end{center}
\vskip  2mm

We write,  using the definitions given in II,  the  amplitudes for
the decays $D^0 \rightarrow K\rho$ as
\begin{eqnarray}
A(D^0 \rightarrow K^{-} \rho^+ )&=&2 \tilde{G}_F  f_\rho m_\rho
(\varepsilon^*.p_D)   F_1^{DK}(m_\rho^2)   (a_1 ^{eff})_{K \rho} \; ,
\nonumber \\
A(D^0 \rightarrow \bar{K}^0 \rho^0 )&=&\sqrt{2} \tilde{G}_F  f_K
m_\rho (\varepsilon^*.p_D) A_0^{D\rho}(m_K^2)  (a_2 ^{eff})_{K \rho}
\; ,  \nonumber \\
\hbox{and} \qquad A(D^+ \rightarrow \bar{K}^0 \rho^+ )&=&A(D^0
\rightarrow K^-\rho^+ ) + \sqrt{2}A(D^0 \rightarrow \bar{K}^0 \rho^0
)\;,
\end{eqnarray}
where
\begin{eqnarray}
\left( a_1^{eff} \right)_{K \rho}&=&a_1 \left( 1 +  {F_1^{(1)nf}
\over F_1^{DK}(m_\rho^2)}  + {C_2 \over a_1} {F_1^{(8)nf} \over
F_1^{DK}(m_\rho^2)} \right) \; ,\nonumber \\
\left( a_2^{eff} \right)_{K \rho}&=&a_2 \left( 1 +  {C_1 \over a_2}
{\tilde{A}_0^{(8)nf}\over A_0^{D\rho}(m_K^2)} \right) \; .
\end{eqnarray}
We have also used, in addition to (8) and (9), the following
definitions,
\begin{eqnarray}
\left\langle K^- \rho^+| (\bar{s}c) (\bar{u}d) | D^0
\right\rangle^{nf}&=& 2 \tilde{G}_F  f_\rho  m_\rho F_1^{(1)nf}
(m_\rho^2) (\varepsilon^*.p_D) \; , \nonumber \\
\left\langle K^- \rho^+| H_w^{(8)} | D^0 \right\rangle&=& 2
\tilde{G}_F  f_\rho   m_\rho  F_1^{(8)nf} (m_\rho^2)
(\varepsilon^*.p_D) \; , \nonumber \\
\left\langle \bar{K}^0 \rho^0| \tilde{H}_w^{(8)} | D^0
\right\rangle&=& \sqrt{2} \tilde{G}_F  f_K   m_\rho
\tilde{A}_0^{(8)nf} (m_K^2)  (\varepsilon^*.p_D) \; .
\end{eqnarray}

Fits to $D \rightarrow \bar{K} \rho$ data admit a solution with zero
fsi phases \cite{kp,mark}, thus we assume $ \delta^{K\rho}_{1/2} -
\delta^{K\rho}_{3/2} = 0$. We use  $F_1^{DK}(0)$  from eqn. (29) and,
for want of better information,  the BSW \cite{bsw} value of
$A_0^{D\rho}(0) = 0.67$.   In this decay also we have considered both
monopole (BSWI) and dipole (BSWII)   extrapolations of the form
factors  $F_1^{DK}(q^2)$  and   $A_0^{D\rho}(q^2)$ with $1^-$ pole at
2.11 GeV and $0^-$ pole at 1.87 GeV respectively.  In Table 2 we show
the allowed ranges of the parameter $ \chi_{K\rho}$ and $
\xi_{K\rho}$, defined in the Table, for which the data could be
fitted. The ranges of  effective $a_1$ and $a_2$ in BSWI and BSWII
scenarios are given as follows
\begin{eqnarray}
\hbox{(BSWI)} \qquad 1.17\leq \left( a_1^{eff} \right)_{K \rho} \leq
1.32 \;, \qquad -1.00 \leq \left(a_2^{eff} \right)_{K \rho} \leq
-0.86 \nonumber \\
\hbox{(BSWII)} \qquad 1.02\leq \left( a_1^{eff} \right)_{K \rho} \leq
1.15\;, \qquad -0.92 \leq \left(a_2^{eff} \right)_{K \rho} \leq-0.80
\end{eqnarray}
A discussion of $\chi_{K^* \pi}$ and $\xi_{K^*\pi}$ is given in the
last Section.

\vskip  5mm
\begin{center}
{\bf C. $D^0 \rightarrow K^{-} a_1^+ \; , \; \bar{K}^{0} a_1^0 $ and
$D^+ \rightarrow \bar{K}^{0} a_1^+ $ }
\end{center}
\vskip  2mm

We write, using definitions given in II,  decay amplitudes for $D
\rightarrow K a_1$ as follows,
\begin{eqnarray}
A(D^0 \rightarrow K^- a_1^+)&=&2 \tilde{G}_F f_{a_1} m_{a_1}
(\varepsilon^*.p_D) F_1^{DK}(m^2_{a_1}) (a_1^{eff})_{K a_1} \; ,
\nonumber\\
A(D^0 \rightarrow \bar{K}^0 a_1^0)&=&\sqrt{2} \tilde{G}_F f_{K}
m_{a_1} (\varepsilon^*.p_D) C_1 \tilde{V}_0^{(8)nf}  \; , \nonumber
\\
A(D^+ \rightarrow \bar{K}^0 a_1^+)&=&A(D^0 \rightarrow K^- a_1^+)   +
\sqrt{2} A(D^0 \rightarrow \bar{K}^0 a_1^0) \;,
\end{eqnarray}
where
\begin{eqnarray}
\left( a_1^{eff} \right)_{K a_1} = a_1 \left( 1 + {F_1^{(1)nf} \over
F_1^{DK}(m_{a_1}^2)} + {C_2 \over a_1} {F_1^{(8)nf} \over
F_1^{DK}(m_{a_1}^2)} \right) \; .
\end{eqnarray}
In deriving (44), in addition to (8) and (12), we have used the
following definitions,
\begin{eqnarray}
\left\langle K^- a_1^+|(\bar{s}c)(\bar{u}d)|D^0
\right\rangle^{nf}&=&2 \tilde{G}_F f_{a_1} m_{a_1}(\varepsilon^*.p_D)
F^{(1)nf} \;, \nonumber \\
\left\langle K^- a_1^+|H_w^{(8)}|D^0 \right\rangle^{nf}&=&2
\tilde{G}_F f_{a_1} m_{a_1}(\varepsilon^*.p_D) F^{(8)nf} \;,\nonumber
\\
\left\langle \bar{K}^0 a_1^0|\tilde{H}_w^{(8)}|D^0
\right\rangle^{nf}&=&\sqrt{2} \tilde{G}_F f_K
m_{a_1}(\varepsilon^*.p_D) \tilde{V}_0^{(8)nf}\;.
\end{eqnarray}

In the decay amplitude for $D^0 \rightarrow \bar{K}^0 a_1^0$ we have
retained only the nonfactorized contribution arising from
$\tilde{H}^{(8)}_w$. The reason being that the factorized amplitude
cannot be calculated in the BSW scheme, $a_1(1260)$ being a $^3P_1$
state, unlike for $K^*$ which is a $^3S_1$ state, BSW procedure does
not define the null-plane wave function for L =1 quark-antiquark
pairs. However, the relevant form factor $V_0^{Da_1}(q^2)$ (see
eq.(12)) can be calculated in the model proposed by Isgur, Scora,
Grinstein and Wise  \cite{isgw} where it can be shown that it
vanishes at the zero-recoil point. This does not imply that it
vanishes everywhere but as it also comes multiplied by the rather
small coefficient $a_2 (\approx -0.09)$, we have neglected the
factorized amplitude all together.  In contrast,  the nonfactorized
term contributing to color-suppressed decay $D^0 \rightarrow
\bar{K}^0 a_1^0$ which we retain in (44), is multiplied by a
relatively large Wilson coefficient $C_1$.

We use  $F_1^{DK}(0)$  from eqn. (29) and  both monopole (BSWI) and
dipole (BSWII)  forms  for $q^2$ extrapolation of the form factor
$F_1^{DK}(q^2)$   with $1^-$ pole at 2.11 GeV.   The results are
given in Table 3. The allowed range of effective $a_1$ is given as
follows,
\begin{eqnarray}
\hbox{(BSWI)} \qquad 2.28 \leq \left(a_1^{eff} \right)_{K a_1} \leq
2.66\;,\nonumber \\
\hbox{(BSWII)} \qquad 1.51 \leq \left(a_1^{eff} \right)_{K a_1} \leq
1.76\;.
\end{eqnarray}

As $a_1 = 1.09$, it may be concluded from (47) that there are large
nonfactorized contributions in $D \rightarrow \bar{K} a_1$ decays.
This is not unanticipated  as the final state particles are
relatively slow in this process.

\vskip  5mm
\begin{center}
{\bf V.  $D \rightarrow  V_1 V_2$}\\
\vskip  2mm
\noindent {\bf A. $D^0 \rightarrow K^{*-} \rho^+ \; , \; \bar{K}^{*0}
\rho^0 $ and  $D^+ \rightarrow \bar{K}^{*0} \rho^+ $ }
\end{center}
\vskip  2mm

Using the definitions given in Section II  one can write the decay
amplitudes for $D^0 \rightarrow K^{*-} \rho^+$,  $\bar{K}^{*0}
\rho^0$ and $D^+ \rightarrow \bar{K}^{*0} \rho^+$ as follows,
\begin{eqnarray}
A(D^0 \rightarrow K^{*-} \rho^+)&=&2 \tilde{G}_Fm_\rho f_\rho \left\{
(m_D + m_{K^*}) \varepsilon_{K^*}.\varepsilon_\rho (a_1
A_1^{DK^*}(m_\rho^2) + a_1 A_1^{(1)nf} + C_2 A_1^{(8)nf} ) \right.
\nonumber \\
 &-& \left. {\varepsilon_{K^*}.(p_D - p_{K^*}) \varepsilon_\rho.(p_D
+ p_{K^*}) \over m_D + m_{K^*}}  a_1 A_2^{DK^*}(m_\rho^2) \right.
\nonumber \\
&+&\left. {2i\over m_D + m_{K^*}}  \varepsilon_{\mu\nu\sigma\delta}
\varepsilon^\mu_\rho \varepsilon^\nu_{K^*} p^\sigma_{K^*} p^\delta_D
a_1 V^{DK^*}(m_\rho^2)  \right\} \; ,\nonumber \\
A(D^0 \rightarrow \bar{K}^{*0} \rho^0)&=&\sqrt{2} \tilde{G}_Fm_{K^*}
f_{K^*} \left\{  (m_D + m_\rho) \varepsilon_{K^*}.\varepsilon_\rho
(a_2 A_1^{D\rho}(m_{K^*}^2) + C_1 \tilde{A}_1^{(8)nf} ) \right.
\nonumber \\
 &-& \left. {\varepsilon_{\rho}.(p_D - p_\rho) \varepsilon_{K^*}.(p_D
+ p_\rho) \over m_D + m_\rho} a_2 A_2^{D\rho}(m_{K^*}^2)   \right.
\nonumber \\
&+&\left. {2i\over m_D + m_\rho}  \varepsilon_{\mu\nu\sigma\delta}
\varepsilon^\mu_{K^*} \varepsilon^\nu_{\rho} p^\sigma_{\rho}
p^\delta_D  a_2 V^{D\rho}(m_{K^*}^2)   \right\} \; , \nonumber \\
\hbox{and} \;\: A(D^+ \rightarrow \bar{K}^{*0} \rho^+)&=&A(D^0
\rightarrow K^{*-} \rho^+) + \sqrt{2} A(D^0 \rightarrow \bar{K}^{*0}
\rho^0)\;,
\end{eqnarray}
where the quantities with super index 1 (e.g. $A_1^{(1)nf}$) arise
from the nonfactorized contribution to the matrix elements of the
color-singlet currents $(\bar{s}c)(\bar{u}d)$; those with super index
8 (e.g. $A_1^{(8)nf}$) arise from $H_w^{(8)}$ made up of color-octet
currents and tildaed quantities (e.g. $\tilde{A}_1^{(8)nf}$) arise
from $\tilde{H}_w^{(8)}$. In writing (48) we have retained
nonfactorized contribution only to S-waves, i.e. $A_1^{(1)nf}$,
$A_1^{(8)nf}$ and $\tilde{A}_1^{(8)nf}$, and neglected all other
nonfactorized contributions as we did in  \cite{ks1} and \cite{ks2}.

The decay rate can then be calculated using (14). For the form
factors we use the following normalizations (only central values of
the experimental numbers are used),
\begin{eqnarray}
&&A_1^{DK^*}(0) = 0.61\pm 0.05 \;, \;\;A_2^{DK^*}(0) = 0.45\pm 0.09
\;, \;\;V^{DK^*}(0) = 1.16\pm 0.16  \;\; \cite{wthrl} \; ,\nonumber
\\
&&\;\;A_1^{D\rho}(0) = 0.78 \;, \qquad \qquad \; A_2^{D\rho}(0) =
0.92 \;, \qquad \qquad \; V^{D\rho}(0) = 1.23 \;\; \cite{bsw} \;,
\end{eqnarray}
and extrapolate them to relevant $q^2$ with monopole forms with pole
masses 2.53 GeV for $A_1^{DK^*}$ and $A_2^{DK^*}$, 2.11 GeV for
$V^{DK^*}$, 2.42 GeV for $A_1^{D\rho}$ and $A_2^{D\rho}$ and 2.01 GeV
for $V^{D\rho}$. We account for nonfactorized contributions through
two parameters $\kappa$ and $\tilde{\kappa}$,
\begin{eqnarray}
\kappa \equiv 1 + {C_2 \over a_1} \chi_{K^*\rho}\;, \qquad
\tilde{\kappa} \equiv 1 + {C_1 \over a_2} \xi_{K^*\rho}\;,
\end{eqnarray}
with
\begin{eqnarray}
\chi_{K^*\rho} = {A_1^{(8)nf} \over A_1^{DK^*}(m_\rho^2)} + {a_1
\over C_2} {A_1^{(1)nf} \over A_1^{DK^*}(m_\rho^2)}\;, \qquad
\hbox{and} \qquad  \xi_{K^*\rho} = {\tilde{A}_1^{(8)nf} \over
A_1^{D\rho}(m_{K^*}^2)} \;.
\end{eqnarray}

We find that  agreement of data with the  calculated branching ratios
is possible for $\xi_{ K^*\rho } $  and  $\chi_{ K^*\rho }  $ lying
in the following range,
\begin{eqnarray}
0.02 \leq \chi_{ K^* \rho } \leq 0.80 \;, \qquad -0.31 \leq \xi_{ K^*
\rho } \leq -0.24\;.
\end{eqnarray}

In particular with $\chi_{ K^* \rho  } = 0.41$  and   $\xi_{ K^* \rho
} = -0.28$ we get,

\begin{eqnarray}
B(D^0 \rightarrow K^{*-} \rho^+ ) = 5.61 \%
\;(\hbox{expt.}\cite{pdg}: (5.9 \pm 2.4)\% )\;, \nonumber \\
B(D^0 \rightarrow \bar{K}^{*0} \rho^0 ) = 1.63 \%
\;(\hbox{expt.}\cite{pdg}: (1.6 \pm 0.4)\% )\;, \nonumber \\
B(D^+ \rightarrow \bar{K}^{*0} \rho^+ ) = 1.81 \%
\;(\hbox{expt.}\cite{pdg}: (2.1 \pm 1.4)\% )\;.
\end{eqnarray}

\vskip  2mm
\begin{center}
{\bf VI.  Summary and Conclusions}\\
\end{center}
\vskip  2mm

We have carried out an analysis of those Cabibbo-favored two-body
hadronic decays of $D^0$ and $D^+$ which involve two isospins in the
final state in a formalism that uses $N_c = 3$ and includes
nonfactorized amplitudes. These decays are: $D \rightarrow \bar{K}
\pi$, $ \bar{K}^* \pi$, $\bar{K} \rho$,  $\bar{K}a_1$ and $\bar{K}^*
\rho$. We have included the measured fsi phases in $\bar{K}\pi$ and
$\bar{K}^*\pi$ decays but only in so far as they rotate the isospin
amplitudes without affecting their magnitudes. We have ignored fsi
phases in $\bar{K}^*\rho$ and $\bar{K}a_1$ decays while the relative
phases is known to be consistent with zero in $\bar{K}\rho$ channel.
We have also ignored  annihilation terms and inelastic fsi. The
rationale for the former is that these terms are proportional to
$a_2$ in $D^0$ decays which in our scheme is only $\approx$ -0.09,
while the neglect of the latter is largely due to ignorance of the
parameters to be used in implementing a believable calculation.

{}From the data,  one only determines $(a_1)^{eff}$ and  $(a_2)^{eff}$
which, as we and others \cite{hyc} have shown, are process-dependent.
The next question is: What effects contribute to   $(a_1)^{eff}$ and
$(a_2)^{eff}$ in a scheme that uses $N_c=3$? We have tacitly assumed
that these effects arise from three sources: the nonfactorized matrix
elements of $H^{(8)}_w = {1 \over 2}
\sum_{a}^{}{(\bar{s}\lambda^ac)(\bar{u}\lambda^ad)}$,
$\tilde{H}^{(8)}_w = {1 \over 2}
\sum_{a}^{}{(\bar{s}\lambda^ad)(\bar{u}\lambda^ac)}$ and the
Hamiltonian made up of color-singlet currents $(\bar{s}c)(\bar{u}d)$.
With these assumptions, we have extracted the relative size of the
nonfactorized contribution in each specific channel. We now turn to a
detailed discussion of specific decays.

{}From $D \rightarrow \bar{K}\pi$ decays we have determined the
parameter $\xi_{K\pi}$ of (30) which is proportional to the matrix
element of $\tilde{H}_w^{(8)}$ denoted by $\tilde{F}^{(8)nf}_0$, to
lie in the range $-0.29 \leq \xi_{K\pi}\leq -0.26$. Cheng \cite{hyc}
determines the same parameter to be -0.36. The small difference could
be due to the fact that we include the fsi phases. We also determine
the parameter $\chi_{K\pi}$ of (30) which includes nonfactorized
contributions from $H^{(8)}_w$ and $(\bar{s}c)(\bar{u}d)$, denoted by
$\tilde{F}_0^{(8)nf}$ and $\tilde{F}_0^{(1)nf}$ respectively in (30).
It would be tempting to assume that $F_0^{(8)nf} =
\tilde{F}_0^{(8)nf}$ and hence extract the value of $F_0^{(1)nf}$.
However, such an assumption would be flawed since $H_w^{(8)}$ and
$\tilde{H}_w^{(8)}$ are related by V-spin symmetry (s
$\Longleftrightarrow $u), but under the same transformation
$|D^0\rangle \rightarrow | D^+_s\rangle$ and  $|K^- \pi^+\rangle
\rightarrow | K^+ \bar{K}^0\rangle$. Thus V-spin symmetry leads to

\begin{eqnarray}
\left\langle K^- \pi^+|H_w^{(8)}|D^0 \right\rangle = \left\langle
K^+\bar{K}^0|\tilde{H}_w^{(8)}|D_s^+ \right\rangle
\end{eqnarray}
 and not to a relation between  $F_0^{(8)nf}$ and
$\tilde{F}_0^{(8)nf}$. Thus these parameters remain unrelated and
very little can be concluded about the size of the nonfactorized
contribution $F_0^{(1)nf}$.  All that can be said is that the
nonfactorized amplitude in $D^0 \rightarrow K^- \pi^+$ decay is
relatively smaller than in $D^0 \rightarrow \bar{K}^0 \pi^0$ decay.
We also emphasize that the nonfactorized contribution in the
color-suppressed decay $D^0 \rightarrow \bar{K}^0 \pi^0$ is enhanced
relative to the factorized term by a factor of $C_1/a_2 (\approx
-14)$ which is not the case in the color-favored decay $D^0
\rightarrow K^- \pi^+$.  Thus the color-suppressed processes are more
likely to reveal the presence of nonfactorized contributions than
color-favored processes. Further,  in the color-favored decay the
nonfactorized amplitude  arising from the color-singlet currents
$(\bar{s}c)(\bar{u}d)$ (called $F_0^{(1)nf}$ here) could be as
important as the one from $H_w^{(8)}$ (called $F_0^{(8)nf}$ here).

In $D \rightarrow \bar{K}^* \pi$ decays we find the parameter
$\xi_{K^* \pi}$, defined in Table 1, to be in the range $-0.34 \leq
\xi_{K^*\pi} \leq -0.28$ for monopole form factors and in the range
$-0.26 \leq \xi_{K^* \pi} \leq -0.21$ for dipole form factors. These
values are considerably smaller than $\approx$ -0.61 given in
\cite{hyc}. The difference could again be due to our inclusion of the
fsi phases which are large. Our estimate of $\xi_{K^*\pi}$ implies
$-0.52 \leq (a_2^{eff})_{K^*\pi} \leq -0.44$ for monopole form
factors and $-0.42 \leq (a_2^{eff})_{K^*\pi} \leq -0.35$ for dipole
form factors. We also find significantly large nonfactorized effects
in the color-favored decay $D^0 \rightarrow K^{*-} \pi^+$ resulting
in $1.77 \leq (a_1^{eff})_{K^* \pi} \leq 1.92$. We recall that it had
been shown in \cite{kp} that factorization assumption together with
$a_1 = 1.26$ and $a_2 = -0.51$ had underestimated $B(D^0 \rightarrow
K^{*-}\pi^+) + B(D^0 \rightarrow \bar{K}^{*0}\pi^0)$ and $B(D^+
\rightarrow \bar{K}^{*0}\pi^+)$. It is the larger value of
$(a_1^{eff})_{K^* \pi}$ that allows a resolution of the problem.

In the color-suppressed decay $D^0 \rightarrow \bar{K}^0 \rho^0$, we
find large nonfactorized contributions: $-0.72 \leq \xi_{K\rho} \leq
-0.61$ for monopole form factor and $-0.66 \leq \xi_{K\rho} \leq
-0.56$ for dipole form factors. They result in  $-1.00 \leq
(a_2^{eff})_{K\rho} \leq -0.86$ and   $-0.92 \leq
(a_2^{eff})_{K\rho} \leq -0.80$ respectively. These parameters help
resolve the problem with factorization assumption which predicted
\cite{kp} too large  a branching ratio for $D^+ \rightarrow \bar{K}^0
\rho^+$ if $a_1 = 1.26$ and $a_2 = -0.51$ were used. With a much
larger effective $a_2$ the rate for $D^+ \rightarrow \bar{K}^0
\rho^+$ is brought down to the experimental value due to a larger
destructive interference between effective $a_1$ and $a_2$. The
nonfactorized contribution to the color-favored decay $D^0
\rightarrow K^- \rho^+$ appears to be small leading to:  $1.17 \leq
(a_1^{eff})_{K\rho} \leq 1.32$  for monopole form factors and   $
1.02\leq  (a_1^{eff})_{K\rho} \leq 1.15$  for dipole form factors.

The decays $D \rightarrow \bar{K} a_1$ have long posed a problem for
the  factorization model. Inclusion of nonfactorized amplitudes
allows us to understand the branching ratios involved. Our picture
suggests that the color-suppressed decay $D^0 \rightarrow \bar{K}^0
a_1^0$ proceeds almost entirely through a nonfactorized amplitude
whose size we limit by the experimental upper limit on $B(D^0
\rightarrow \bar{K}^0a_1^0)$. We are then able to understand the
measured branching ratios $B(D^0 \rightarrow K^- a^+_1)$  and  $B(D^+
\rightarrow \bar{K}^0 a^+_1)$ provided that:   $ 2.28\leq
(a_1^{eff})_{Ka_1} \leq 2.66$ for monopole form factors and  $
1.51\leq  (a_1^{eff})_{Ka_1} \leq 1.76$ for dipole form factors.
These large values of $(a_1^{eff})$ are not unexpected for this mode
where the final-state particles are relatively slow.

For the decays $D \rightarrow K^* \rho $ (and, in general, for any $
P \rightarrow VV $ decay) one cannot define $(a_1^{eff})$ and
$(a_2^{eff})$ as the decay amplitude involves three independent
Lorentz scalar structures and it is not possible to factor out  an
effective $a_1$ and $a_2$. However, retaining the nonfactorized
effects only in S-wave final states, we find significant
nonfactorized effects in the color-suppressed decay $D^0 \rightarrow
\bar{K}^{*0} \rho^0$ characterized by the parameter $\xi_{K^*\rho}$
of eqn. (50): $-0.31 \leq  \xi_{K^* \rho} \leq -0.24$. The analogous
parameter $\chi_{K^* \rho}$, eq.(50), which is a measure of
nonfactorized contribution to the color-favored decay $D^0
\rightarrow K^{*-} \rho^+$ has the opposite sign, and could in
principle, be very small:  $0.02 \leq \chi_{K^*\rho} \leq 0.80$.

We conclude by saying that one can understand D decays in a picture
with $N_c = 3$ but with the inclusion of nonfactorized amplitudes.
This picture results in process-dependent  effective $a_1$ and $a_2$,
which ought to be complex as are all the nonfactorized amplitudes. We
have not included the inelastic final-state interaction effects which
would further complicate the analysis. The effort here was to
parametrize the nonfactorized amplitudes and determine their size.
The understanding of any systematics that emerge is yet to come.

\begin{center}
{\bf  Acknowledgements}
\end{center}

ANK wishes to acknowledge a research grant from the Natural Sciences
and Engineering Research Council of Canada which partially supported
this research.

\newpage

%--------------------------- K^* \pi -----------------------

\begin{table}
\begin{center}
\caption{ $D \rightarrow K^* \pi $ }
\vskip 3mm
\begin{tabular}{|c|c|c|c|c|}
\hline
&\multicolumn{3}{c|}{Scenario1 (BSWI)} &\\ \cline {2 -4}
\multicolumn{1}{|c}{Channel}&
\multicolumn{1}{|c|}{$\chi_{K^*\pi} $}&
\multicolumn{1}{|c|}{$\xi_{K^*\pi} $}&
\multicolumn{1}{|c|}{Th.BR$^{(a)}$}&
\multicolumn{1}{c|}{Expt.BR$^{(c)}$} \\
\multicolumn{1}{|c}{}&
\multicolumn{1}{|c|}{ }&
\multicolumn{1}{|c|}{}&
\multicolumn{1}{|c|}{in $\%$}&
\multicolumn{1}{c|}{in $\%$} \\
\hline
$D^0 \rightarrow K^{*-} \pi^+$&&&4.63&4.9 $\pm$ 0.6\\
$D^0 \rightarrow \bar{K}^{*0} \pi^0$&-1.62$\leq \chi_{K^*\pi} \leq$
-1.33&-0.34$\leq \xi_{K^*\pi} \leq$ -0.28&3.14&3.0 $\pm$ 0.4\\
$D^+ \rightarrow \bar{K}^{*0} \pi^+$&&&2.07&2.2 $\pm$ 0.4\\
\hline
\end{tabular}

\begin{tabular}{|c|c|c|c|c|}
\hline
&\multicolumn{3}{c|}{Scenario 2 (BSWII)} &\\ \cline {2 -4}
\multicolumn{1}{|c}{Channel}&
\multicolumn{1}{|c|}{$\chi_{K^*\pi} $}&
\multicolumn{1}{|c|}{$\xi_{K^*\pi} $}&
\multicolumn{1}{|c|}{Th.BR$^{(b)}$}&
\multicolumn{1}{c|}{Expt.BR$^{(c)}$} \\
\multicolumn{1}{|c}{}&
\multicolumn{1}{|c|}{}&
\multicolumn{1}{|c|}{}&
\multicolumn{1}{|c|}{in $\%$}&
\multicolumn{1}{c|}{in $\%$} \\
\hline
$D^0 \rightarrow K^{*-} \pi^+$&&&4.62&4.9 $\pm$ 0.6\\
$D^0 \rightarrow \bar{K}^{*0} \pi^0$&-1.60$\leq \chi_{K^*\pi} \leq$
-1.34&-0.26$\leq \xi_{K^*\pi} \leq$ -0.21&3.17&3.0 $\pm$ 0.4\\
$D^+ \rightarrow \bar{K}^{*0} \pi^+$&&&2.24&2.2 $\pm$ 0.4\\
\hline
\end{tabular}
\end{center}
$\chi_{K^*\pi}  \equiv {A_0^{(8)nf} \over A_0^{DK^*}(m_\pi^2)} + {a_1
\over C_2} {A_0^{(1)nf} \over A_0^{DK^*}(m_\pi^2)} $\\
$\xi_{K^*\pi}  \equiv {\tilde{F}_1^{(8)nf} \over
F_1^{D\pi}(m_{\bar{K^*}}^2)}$\\
(a) for $\chi_{K^*\pi}  = -1.47$ and $\xi_{K^*\pi}  = -0.31$ \\
(b) for $\chi_{K^*\pi}  = -1.47$ and $\xi_{K^*\pi}  = -0.23$ \\
(c) Source Ref. \cite{pdg}
\end{table}

%--------------------------- K \rho -----------------------

\begin{table}
\begin{center}
\caption{ $D \rightarrow K \rho $ }
\vskip 3mm
\begin{tabular}{|c|c|c|c|c|}
\hline
&\multicolumn{3}{c|}{Scenario 1 (BSWI)} &\\ \cline {2 -4}
\multicolumn{1}{|c}{Channel}&
\multicolumn{1}{|c|}{$\chi_{K\rho} $}&
\multicolumn{1}{|c|}{$\xi_{K\rho} $}&
\multicolumn{1}{|c|}{Th.BR$^{(a)}$}&
\multicolumn{1}{c|}{Expt.BR$^{(c)}$} \\
\multicolumn{1}{|c}{}&
\multicolumn{1}{|c|}{}&
\multicolumn{1}{|c|}{}&
\multicolumn{1}{|c|}{in $\%$}&
\multicolumn{1}{c|}{in $\%$} \\
\hline
$D^0 \rightarrow K^- \rho^+$&&&10.32&10.40 $\pm$ 1.30\\
$D^0 \rightarrow \bar{K}^0 \rho^0$&-0.45$\leq \chi_{K\rho} \leq$
-0.16&-0.72$\leq \xi_{K\rho} \leq$ -0.61&1.10&1.10 $\pm$ 0.18\\
$D^+ \rightarrow \bar{K}^0 \rho^+$&&&7.56&6.60 $\pm$ 2.50\\
\hline
\end{tabular}

\begin{tabular}{|c|c|c|c|c|}
\hline
&\multicolumn{3}{c|}{Scenario 2 (BSWII)} &\\ \cline {2 -4}
\multicolumn{1}{|c}{Channel}&
\multicolumn{1}{|c|}{$\chi_{K\rho} $}&
\multicolumn{1}{|c|}{$\xi_{K\rho} $}&
\multicolumn{1}{|c|}{Th.BR$^{(b)}$}&
\multicolumn{1}{c|}{Expt.BR$^{(c)}$} \\
\multicolumn{1}{|c}{}&
\multicolumn{1}{|c|}{}&
\multicolumn{1}{|c|}{}&
\multicolumn{1}{|c|}{in $\%$}&
\multicolumn{1}{c|}{in $\%$} \\
\hline
$D^0 \rightarrow K^- \rho^+$&&&10.36&10.40 $\pm$ 1.30\\
$D^0 \rightarrow \bar{K}^0 \rho^0$&-0.11$\leq \chi_{K\rho}
\leq$\,\,\,\,0.14&-0.66$\leq \xi_{K\rho} \leq$ -0.56&1.08&1.10 $\pm$
0.18\\
$D^+ \rightarrow \bar{K}^0 \rho^+$&&&7.76&6.60 $\pm$ 2.50\\
\hline
\end{tabular}
\end{center}
$\chi_{K\rho}  \equiv {F_1^{(8)nf} \over F_1^{DK}(m_\rho^2)} + {a_1
\over C_2} {F_1^{(1)nf} \over F_1^{DK}(m_\rho^2)}  $   \\
$\xi_{K\rho}  \equiv {\tilde{A}_0^{(8)nf} \over
A_0^{D\rho}(m_{\bar{K}}^2)}$\\
(a) for $\chi_{K\rho}  = -0.30$ and $\xi_{K\rho}  = -0.67$ \\
(b) for $\chi_{K\rho}  = 0.02$ and $\xi_{K\rho}  = -0.61$ \\
(c) Source Ref. \cite{pdg}
\end{table}

%........................ K a_1...................

\begin{table}
\begin{center}
\caption{ $D \rightarrow K a_1 $ }
\vskip 3mm
\begin{tabular}{|c|c|c|c|c|}
\hline
&\multicolumn{3}{c|}{Scenario 1 (BSWI)} &\\ \cline {2 -4}
\multicolumn{1}{|c}{Channel}&
\multicolumn{1}{|c|}{}&
\multicolumn{1}{|c|}{}&
\multicolumn{1}{|c|}{Th.BR$^{(a)}$}&
\multicolumn{1}{c|}{Expt.BR$^{(c)}$} \\
\multicolumn{1}{|c}{}&
\multicolumn{1}{|c|}{$\chi_{Ka_1} $}&
\multicolumn{1}{|c|}{$\tilde{V}_0^{(8)nf} $}&
\multicolumn{1}{|c|}{in $\%$}&
\multicolumn{1}{c|}{in $\%$} \\
\hline
$D^0 \rightarrow K^- a_1^+$&&&7.87&7.9 $\pm$ 1.2\\
$D^0 \rightarrow \bar{K}^0 a_1^0$&-3.07$\leq \chi_{Ka_1} \leq$
-2.34&-1.54$\leq \tilde{V}_0^{(8)nf}  \leq$ -0.89&0.62& $<$ 1.9\\
$D^+ \rightarrow \bar{K}^0 a_1^+$&&&7.19&8.1 $\pm$ 1.7\\
\hline
\end{tabular}

\begin{tabular}{|c|c|c|c|c|}
\hline
&\multicolumn{3}{c|}{Scenario 2 (BSWII)} &\\ \cline {2 -4}
\multicolumn{1}{|c}{Channel}&
\multicolumn{1}{|c|}{}&
\multicolumn{1}{|c|}{}&
\multicolumn{1}{|c|}{Th.BR$^{(b)}$}&
\multicolumn{1}{c|}{Expt.BR$^{(c)}$} \\
\multicolumn{1}{|c}{}&
\multicolumn{1}{|c|}{$\chi_{Ka_1} $}&
\multicolumn{1}{|c|}{$\tilde{V}_0^{(8)nf} $}&
\multicolumn{1}{|c|}{in $\%$}&
\multicolumn{1}{c|}{in $\%$} \\
\hline
$D^0 \rightarrow K^- a_1^+$&&&7.90&7.9 $\pm$ 1.2\\
$D^0 \rightarrow \bar{K}^0 a_1^0$&-1.30$\leq \chi_{Ka_1} \leq$
-0.83&-1.55$\leq \tilde{V}_0^{(8)nf}  \leq$ -0.69&0.53& $<$ 1.9\\
$D^+ \rightarrow \bar{K}^0 a_1^+$&&&8.06&8.1 $\pm$ 1.7\\
\hline
\end{tabular}
\end{center}
$\chi_{Ka_1}  \equiv {F_1^{(8)nf} \over F_1^{DK}(m_{a_1}^2)} + {a_1
\over C_2} {F_1^{(1)nf} \over F_1^{DK}(m_{a_1}^2)} $\\
(a) for $\chi_{Ka_1}  = -2.71$ and $\tilde{V}_0^{(8)nf}   = -1.22$\\
(b) for $\chi_{Ka_1}  = -1.07$ and $\tilde{V}_0^{(8)nf}   = -1.12$ \\
(c) Source Ref. \cite{pdg}
\end{table}

\end{document}